\documentclass[10pt,journal,compsoc]{IEEEtran}
\ifCLASSOPTIONcompsoc
  \usepackage{cite}
\else
  \usepackage{cite}
\fi
\ifCLASSINFOpdf
  \usepackage[pdftex]{graphicx}
\else
  \usepackage[dvips]{graphicx}
\fi
\usepackage{amsmath}
\usepackage{amssymb}

\usepackage{algorithmic}

\usepackage{array}

\ifCLASSOPTIONcompsoc
 \usepackage[caption=false,font=footnotesize,labelfont=sf,textfont=sf]{subfig}
\else
 \usepackage[caption=false,font=footnotesize]{subfig}
\fi
\usepackage{fixltx2e}
\usepackage{url}

\begin{document}
%
\title{Music Sequence Prediction with Mixture Hidden Markov Models}



%
\author{Tao~Li,
Minsoo~Choi,
Kaiming~Fu,
Lei~Lin
\IEEEcompsocitemizethanks{\IEEEcompsocthanksitem Tao Li is with the Department
of Computer Science, Purdue University, West Lafayette, IN 47907, USA. \protect\\
E-mail: taoli@purdue.edu
\IEEEcompsocthanksitem Minsoo Choi is with the School of Industrial Engineering, Purdue University, West Lafayette, IN 47907, USA.
\IEEEcompsocthanksitem Kaiming Fu is with the School of Mechanical Engineering, Purdue University, West Lafayette, IN 47907, USA.
\IEEEcompsocthanksitem Lei Lin is with the Goergen Institute for Data Science, University of Rochester, Rochester, NY 14627, USA.
}
\thanks{Accepted to the 4th International Conference on Artificial Intelligence and Applications (AI 2018), October 27-28, 2018, Dubai, UAE.}
}

\IEEEtitleabstractindextext{%
\begin{abstract}
Recommendation systems that automatically generate personalized music playlists for users have attracted tremendous attention in recent years. Nowadays, most music recommendation systems rely on item-based or user-based collaborative filtering or content-based approaches. In this paper, we propose a novel mixture hidden Markov model (HMM) for music play sequence prediction. We compare the mixture model with state-of-the-art methods and evaluate the predictions quantitatively and qualitatively on a large-scale real-world dataset in a Kaggle competition. Results show that our model significantly outperforms traditional methods as well as other competitors. We conclude by envisioning a next-generation music recommendation system that integrates our model with recent advances in deep learning, computer vision, and speech techniques, and has promising potential in both academia and industry.
\end{abstract}

}

\maketitle

\IEEEdisplaynontitleabstractindextext

%
\IEEEpeerreviewmaketitle

\IEEEraisesectionheading{\section{Introduction}\label{sec:intro}}
Traditional ways of listening to music, such as playing CDs or tapes, are being replaced by online music stores and streaming services\cite{casey2008content,van2013deep}. A report from {\em the International Federation of the Phonographic Industry}\cite{ifpi2018global} shows that the revenue ofstreaming industry increased by 41.1\% last year, and now the streaming industry is the most significant portion of the music industry\cite{li2018youtube,li2017youtubestat}. As opposed to traditional ways, the streaming services enable users to interact with music platforms dynamically. For example, one can select and delete a song from his/her playlist any time and replace it with an entirely different one. However, it is inconvinient for users to manually update the playlists every time they listen to a new song. To improve users' listening experience, various intelligent personalized music recommendation systems have been proposed\cite{chen2001music,hicken2005music,park2006context,alcalde2006method,song2012survey,hoffmann2014music}.

{\em Downie}\cite{downie2003music} introduced the discipline of music information retrieval (MIR), which is a multidisciplinary research endeavor that retrieves comprehensive data from music, including users' preferences as well as artworks, genres, beats, etc., in order to enhance the music-listening experience. MIR already has many real-world applications, powered by latest technologies ranging from computational musicology, audio analysis, and user preferences mining, to human-computer interaction\cite{bel1993computational,pfeiffer1997automatic,holland2003preference,dix2009human}.

Music sequence prediction is a particular type of recommendation system problem. Unlike usual recommendation system applications (e.g., Amazon and Netflix), predicting music sequences is more challenging and requires application-specific tailoring. Considering how people listen to music in their daily lives, they usually play CDs or their playlists in Spotify repeatedly when they are in a car or at a gym\cite{baltrunas2011incarmusic,wang2018cooperative,li2018modeling,gong2018cooperative}. However, it is unlikely that one would manually select songs or artists. Instead, users tend to repeat the same sequence of music, often in a pre-made playlist, unless one subscribed to a special recommendation service such as music station in Spotify. In this project, we verified these intuitions with some descriptive statistics analysis of training data and by comparing different models, discussed in the comparison section. The one important takeaway of this paper is that it doesn't deal with single factor such as music labels or sequential information separately but instead integrates all related factors together for the recommendation.

The rest of the paper is organized as follows: Section \ref{sec:dataset} provides a review of the dataset. Section \ref{sec:method} discusses state-of-the-art algorithms and our novel mixture model. Experiments are shown in Section \ref{sec:experiment}. In Section \ref{sec:future}, we outline limitations and provide future recommendations. Section \ref{sec:conclusion} concludes the paper.

\section{The Dataset}\label{sec:dataset}
We use the dataset provided by the {\em Kaggle Prediction Competition: Music Sequence Prediction}\cite{kaggle2018}, which consists of artists played by different users on an online streaming service. In the dataset, each user has a sequence of $29$ artists which they have played in order. Specifically, the dataset contains $972$ rows and $29$ columns where each row stands for each user and each column has an identification code of an artist. The order of columns corresponds to the order of music listening sequence of users. For example, if a user has artist $id_1$, $id_2$, and $id_3$ as values for column $c_1$, $c_2$, and $c_3$ respectively, it means that the user listens to music by the artist $id_1$ at first, and then $id_2$ and $id_3$ in his or her listening sequence. Note that the dataset only has information of artists but not of their songs.
In the same way that one could have many songs of the same artist in their playlist, a sequence can have the same id more than once, and we do not differentiate the values, artist's id, no matter where they appear; that is, there is no difference between $artist_1$ appearing on the first column, $c_1$ on a playlist and $artist_1$ appearing on the tenth column $c_{10}$ on the playlist. Lastly, the total number of artists in the pool is $3924$.

The goal of the project is to predict the $30^{th}$ artist that users will play, given the previous $29$ playlists of their music listening sequence. We will give ten candidates of artists for the $30^{th}$, where the order of candidates has meaning regarding prediction accuracy. We assume that the earlier an artist appears, the more probable it is that the user plays the artist's song. It is reasonable to give a sequence of artists as the predictor instead of one target artist regarding that our ultimate objective is on recommendation not on exact prediction itself. For recommending purposes, figuring out preference tendencies of users would be more meaningful than just acquiring one prediction for the $30^{th}$ artist they play. Accordingly, we used an evaluation metric which is devised to evaluate this sequential information of the predictors.

\section{Methods}\label{sec:method}
The problem of music sequence prediction is intriguing because its a combination of techniques in both recommendation system and sequence prediction, which have been studied extensively. In this section, we first discuss those state-of-the-art methods and then introduce our approach to attack the new issue.

\subsection{User-based Collaborative Filtering}
Collaborative filtering (CF) is a machine learning algorithm for a user-specific recommendation system, which does not need extrageneous information about users or items. Instead, CF utilizes internal user-item information which represents preferences for items by users. \cite{koren2015advances} Depending on the system, the user-item information can vary from ratings to purchase history. CF uses the information to find neighborhood and utilize it to predict a specific user’s preference or future behavior. Although there are many variants, CF-based algorithms fall into categories: user-based and item-based\cite{ekstrand2011collaborative}.

User-based CF assumes that, if some users whose past rating behaviors on other items are similar to user $u$, $u$ tends to behave similarly to an item $i$.
User-based CF, therefore, uses the similarity function $s$ with respect to user dimension. $s: U\times U\rightarrow \mathbb{R}$. There are various kinds of similarity functions such as Pearson correlation and cosine similarity.
Using this similarity between the user u and others, user-based CF computes a neighborhood which consists of the most similar k users. Then it computes the prediction of the user $u$'s preference to item $i$ combining the ratings of the neighborhood. How many neighbors to select also depends on the problem.
\begin{equation}
	P_{u,i} = \bar{r}_u + \frac{\sum_{u' \in N}s(u, u')(r_{u',i} - \bar{r}_{u'})}{\sum_{u' \in N} |s(u, u')|}
\end{equation}
where $\bar{r}_u$ is the average rating of the user $u$, $r_{u,i}$ is a rating of item $i$ by user $u$, and $s(I, j)$ is the similarity score between user $i$ and user $j$.

\subsection{Item-based Collaborative Filtering}
While the user-based CF algorithm is intuitive and effective, it is computationally expensive and has problems with handling sparsity. The most important reason is because the algorithm does and has to calculate similarity scores at the time when predictions or recommendations are needed; each operation takes $O(U)$. In a system which has large number of users and fast changing such as e-commerce, this cost could be prohibiting. Item-based CF solves the problem allowing the system to pre-compute and store the similarity scores and use the pre-stored data at the time when it predicts or recommends items to users.
Item-based CF makes pre-computing possible using similarities between the rating patterns of items instead of users. The methods for computing item similarity are the same as user-based CF, except that the item-based CF computes with respect to items. The most popular similarity metrics is cosine similarity.
\begin{equation}
	s(u, v) = \frac{\mathbf{r}_u \cdot \mathbf{r}_v}{||\mathbf{r}_u||_2 \cdot ||\mathbf{r}_v||_2}
\end{equation}
After collecting a similarity scores of items similar to $i$, item-based CF finds the most similar k neighbors as user-based CF. Then, it predicts user $u$'s rating with respect to the neighborhood as follows:
\begin{equation}
	P_{u,i} = \bar{r}_u + \frac{\sum_{u'\in N}s(u,u')(r_{u',i} - \bar{r}_{u'})}{\sum_{u'\in N} |s(u, u')| }
\end{equation}
For non-real-values ratings scales, such as purchase history data which can be represented as binary, we can compute pseudo-predictions with some simple tactics such as summation of all k similarities.

\subsection{Hidden Markov Models}
The Hidden Markov Model (HMM) is an extended version of the Markov Chain Model which is a stochastic process following Markovian property. Markovian property is an assumption on which the future state depends only on a current state which can be written as \cite{gagniuc2017markov}
\begin{equation}
    P(q_t | q_{t-1}, q_{t-2}, \dots , q_0)=P(q_t | q_{t-1})
\end{equation}
However, the Markov Model has a limitation in that it assumes each state corresponds to an observable event. To overcome the problem, HMM introduces a separation between observations and states of the system. That is, HMM assumes there is a latent and unobservable stochastic process within the system itself, but we can only observe another set of a stochastic processes as observations. With them, we could guess the hidden model of the system \cite{rabiner1989tutorial}. To be specific, we get the hidden model with the following process; decide model topology of the system such as the number of states in the model and the number of distinct observations; set parameters for the transition probability (A={a\textsubscript{ij}}), the observation symbol probability (B={b\textsubscript{j}(K)}), and initial state distribution $\pi$ ={ $\pi$\textsubscript{i}}) which can be written as the following equations:
\begin{equation}
    a_{ij}=P[q_{t+1}=S_i\mid q_t=S_j ], 1 \le i,j \le N
\end{equation}
\begin{equation}
    b_j(k)=P[v_k\, at\, t\mid q_t = S_j ], 1 \le j \le N, 1 \le k \le M
\end{equation}
\begin{equation}
    \pi_i = P[q_1 = S_i], 1 \le i \le N
\end{equation}
Next, with the initial parameters $\lambda=(A, B, \pi)$, we are going to keep re-estimating until we have no or little improvement regarding the probability of the observations given the parameters $\lambda$: $P[O|\lambda]$. The HMM has been applied to various kinds of sequential analysis from natural language processing to genetic sequence alignment \cite{rabiner1989tutorial, hughey1996hidden}.

\subsection{Mixture Models}
The proposed mixture hidden Markov model (MHMM) is given as follow:
\begin{equation}
    \text{MHMM}(n : \mathbb{D}) = \text{HMM}(n_1 : \mathbb{D}) \odot \text{CF}(n_2 : \mathbb{D})
\end{equation}
where $\mathbb{D}$ are observed data, $n_1$ and $n_2$ are hyperparameters subject to $n = n_1 + n_2$, and $\odot$ is a special operator we defined to concatenate two sequences. In practice, there are lots of tied rankings given by HMM and CF. A trick here is that we add weights to each artist, which are learned from the entire artists population - the more frequent a artist in the population, the higher weight he or she has. This trick fixs tied situations and is critical to overall performance which will be shown in section \ref{sec:experiment}.

\section{Experiments}\label{sec:experiment}

\subsection{Evaluation Method}
We use mean average precision at K (MAP@K)\cite{kan2016map} to evaluate the prediction power of models. MAP@K is the average value of precision at K (AP@K), where AP@K is a score which gives higher value for a sequence which has actual value at smaller index\cite{manning2008chapter}. For example, if actual value is $1$ and two prediction sequences are $\{1, 2, 3, 4, 5\}$ and $\{2, 1, 3, 4, 5\}$, the first prediction has higher AP@K score than the second prediction since $1$ comes earlier in the first prediction. The formula of MAP@K for this problem is
\begin{equation}
\text{MAP@K} = \frac{1}{N} \sum_{i=1}^{N} \frac{1}{f(y_i, \mathbb{D}_i)}
\end{equation}
where $N$ is the total number of users, $y_i$ is the actual artist that user $i$ listens to, $\mathbb{D}_i$ is the predicted sequence of size $K$, and
\begin{align*}
f(y_i, \mathbb{D}_i) = \begin{cases}
    \text{index of\ } y_i \text{\ in\ } \mathbb{D}_i,& y_i \in \mathbb{D}_i \\
    0,& y_i \not\in \mathbb{D}_i
\end{cases}
\end{align*}
In our experiments, $N$ is $972$,  $K$ is $10$, and $y_i$ is the $30^\text{th}$ artist that user $i$ listens to.

\subsection{Comparison}
\begin{table}[!t]
\caption{Comparison of Different Models}
\label{tab:compare}
\centering
\begin{tabular}{ | l | r | }
\hline
Model & MAP@K \\
\hline
$HF_{corpus}$ & $0.00721$ \\
$HF_{current}$ & $0.10050$ \\
$CF_{user}$ & $0.01143$ \\
$CF_{item}$ & $0.01233$ \\
$HMM$ & $0.12838$ \\
$MHMM$ & $0.13958$ \\
\hline
\end{tabular}
\end{table}

\begin{figure}[!t]
\centering
\includegraphics[width=2in]{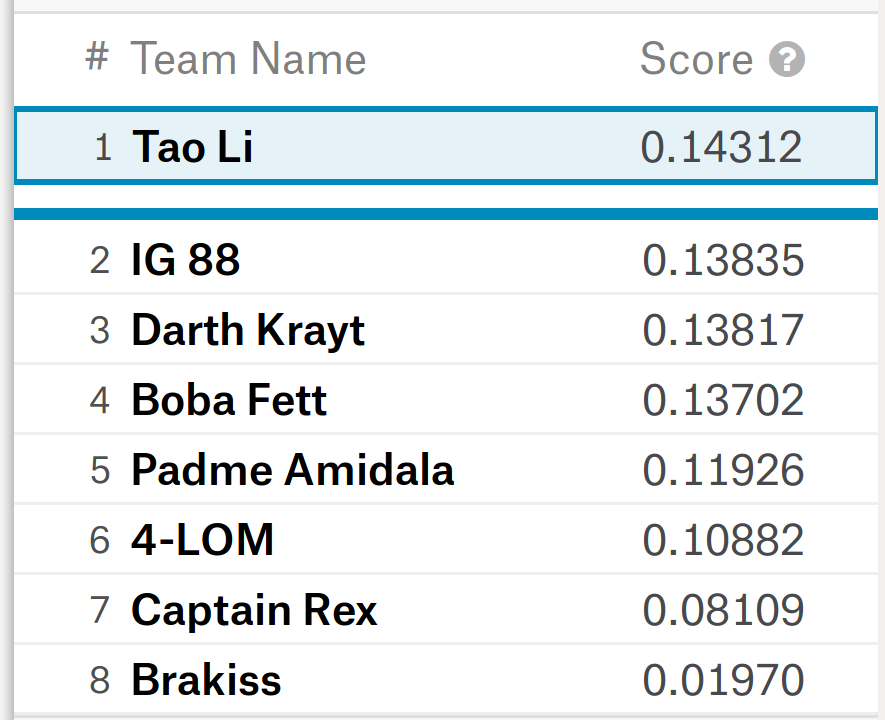}
\caption{Final leaderboard of the competition}
\label{fig:ranking}
\end{figure}

A comparion of different methods can be found in Table \ref{tab:compare}, where $HF_{corpus}$ and $HF_{current}$ mean predicting with the highest frequent artists from the whole corpus and from the current list respectively, $CF_{user}$ and $CF_{item}$ represent user-based collaborative filtering and item-based collaborative filtering respectively, HMM is the hidden Markov model, and MHMM is the mixture model we propose. Results show our method significantly outperform state-of-the-art models. Figure \ref{fig:ranking} is the final leaderboard of the competition.

\section{Limitations and Future Works}\label{sec:future}
Despite the success of our models in the competition and on real-world datasets, there are still many opportunities for further enhancements, including:

\textbf{Similarity Metrics.} For collateral filtering-based models, various cosine similarity-based measurements \cite{li2004similarity} have been proposed and the one used in our models are not necessarily optimal. {\em Resnik}\cite{resnik1999semantic} proposed an information-based measurement, semantic similarity, which might achieve more reasonable similarity results by using customized weights for different users based on information from other sources.

\textbf{Markov Property.} Although the mixture hidden Markov models have convincing performance in the dataset, the markov property might be to strong to assume for general cases. It is natural to come out that using information from previous $n$ steps instead of only one step before to achieve better prediction. {\em Lafferty et al.}\cite{lafferty2001conditional} introduce conditional random fields (CRFs) which takes context into consideration. {\em Sutton et al.}\cite{sutton2006introduction} provide a diagram of the relationship between HMMs and general CRFs as well as naive Bayes, logistic regression, linear-chain CRFs, and generative models.

\textbf{Explore/Exploit Tradeoff.}
A central concern of adaptive learning system is the relation between the exploration of new possibilities and the exploitation of old certainties\cite{march1991exploration}. The explore/exploit (EE) tradeoff, since first introduced\cite{schumpeter1934theory}, it has been discussed extensively in the RecSys community. Although the models we used outperform others in the competition in terms of prediction power, the recommended items are restricted to a limited pool and fail to further explore new possibilities of user preference. As recommendation systems are usually built for commericial purposes, user experience is a major concern. {\em Gaver et al.}\cite{gaver2000alternatives} argued the importance of users' non-instrumental needs, including diversion and surprise, which are needed to be addressed in software systems.

\textbf{Future Techniques.} With the advent and successes of recent deep learning techniques (e.g., CNNs and GANs\cite{lecun2015deep,goodfellow2014generative,lin2018deep,li2018toward}), great improvements have been achieved in computer vision and speech. Advances in these techniques provide new opportunities for music recommendation systems as a system can not only rely on offline information (e.g. music genre, user's playing history) but is also possible to track users' visual and auditory feedbacks in real-time.
Facial expression detection, tracking, and classification of facial expressions have been widely studied\cite{rodriguez2017deep,lopes2017facial,zeng2018facial,li2018single} and even micro-expression detection \cite{li2017towards,takalkar2018survey} for digging hidden emotions is becoming possible.
Many efforts have been made from the music synthesis community\cite{dodge1997computer,wang2014guided,o2017v2hz} and a novel autoencoder framework was recently proposed\cite{colonel2017improving}. We can expect customized real-time music recommendation systems that learn from human, compose for human, and even know human better than human theirselves in the near future.

\section{Conclusion}\label{sec:conclusion}
In this paper, we compare various state-of-the-art methods for music sequence predictions and propose a novel mixture hidden Markov model which shows promising results and significantly outperforms others in a Kaggle competition. We discuss limitations of similarity metrics, Markov-based models and explore/exploit trade-offs. Futhermore, we point out future directions for next-generation music recommendation systems.

\ifCLASSOPTIONcompsoc
  \section*{Acknowledgments}
\else
  \section*{Acknowledgment}
\fi


The authors would like to thank Dr. Bruno Ribeiro for interesting lectures and well-designed projects in {\em Purdue CS 573, Spring 2018}\cite{ribeiro2018cs}. This work was supported by a grant from the Goodata Foundation (GDF grant b5d-8d-564-f0).




\bibliographystyle{IEEEtran}
\bibliography{IEEEabrv,db}

\begin{thebibliography}{10}
\providecommand{\url}[1]{#1}
\csname url@samestyle\endcsname
\providecommand{\newblock}{\relax}
\providecommand{\bibinfo}[2]{#2}
\providecommand{\BIBentrySTDinterwordspacing}{\spaceskip=0pt\relax}
\providecommand{\BIBentryALTinterwordstretchfactor}{4}
\providecommand{\BIBentryALTinterwordspacing}{\spaceskip=\fontdimen2\font plus
\BIBentryALTinterwordstretchfactor\fontdimen3\font minus
  \fontdimen4\font\relax}
\providecommand{\BIBforeignlanguage}[2]{{%
\expandafter\ifx\csname l@#1\endcsname\relax
\typeout{** WARNING: IEEEtran.bst: No hyphenation pattern has been}%
\typeout{** loaded for the language `#1'. Using the pattern for}%
\typeout{** the default language instead.}%
\else
\language=\csname l@#1\endcsname
\fi
#2}}
\providecommand{\BIBdecl}{\relax}
\BIBdecl

\bibitem{casey2008content}
M.~A. Casey, R.~Veltkamp, M.~Goto, M.~Leman, C.~Rhodes, and M.~Slaney,
  ``Content-based music information retrieval: Current directions and future
  challenges,'' \emph{Proceedings of the IEEE}, vol.~96, no.~4, pp. 668--696,
  2008.

\bibitem{van2013deep}
A.~Van~den Oord, S.~Dieleman, and B.~Schrauwen, ``Deep content-based music
  recommendation,'' in \emph{Advances in neural information processing
  systems}, 2013, pp. 2643--2651.

\bibitem{ifpi2018global}
\BIBentryALTinterwordspacing
``Global music report 2018,'' 2018. [Online]. Available:
  \url{http://www.ifpi.org/downloads/GMR2018.pdf}
\BIBentrySTDinterwordspacing

\bibitem{li2018youtube}
T.~Li, L.~Lin, M.~Choi, K.~Fu, S.~Gong, and J.~Wang, ``Youtube av 50k: an
  annotated corpus for comments in autonomous vehicles,'' \emph{arXiv preprint
  arXiv:1807.11227}, 2018.

\bibitem{li2017youtubestat}
\BIBentryALTinterwordspacing
T.~Li, M.~Choi, and L.~Lin, ``Youtubestat.py: a python module to download
  youtube statistics and rankings,'' \emph{GitHub Repository}, pp. 1--3, June
  2017. [Online]. Available:
  \url{https://github.com/Eroica-cpp/YouTube-Statistics}
\BIBentrySTDinterwordspacing

\bibitem{chen2001music}
H.-C. Chen and A.~L. Chen, ``A music recommendation system based on music data
  grouping and user interests,'' in \emph{Proceedings of the tenth
  international conference on Information and knowledge management}.\hskip 1em
  plus 0.5em minus 0.4em\relax ACM, 2001, pp. 231--238.

\bibitem{hicken2005music}
W.~Hicken, F.~Holm, J.~Clune, and M.~Campbell, ``Music recommendation system
  and method,'' Feb.~17 2005, uS Patent App. 10/917,865.

\bibitem{park2006context}
H.-S. Park, J.-O. Yoo, and S.-B. Cho, ``A context-aware music recommendation
  system using fuzzy bayesian networks with utility theory,'' in
  \emph{International conference on Fuzzy systems and knowledge
  discovery}.\hskip 1em plus 0.5em minus 0.4em\relax Springer, 2006, pp.
  970--979.

\bibitem{alcalde2006method}
V.~G. Alcalde, C.~M.~L. Ullod, A.~T. Bonet, A.~T. Llopis, J.~S. Marcos, D.~C.
  Ysern, and D.~Arkwright, ``Method and system for music recommendation,''
  Jul.~25 2006, uS Patent 7,081,579.

\bibitem{song2012survey}
Y.~Song, S.~Dixon, and M.~Pearce, ``A survey of music recommendation systems
  and future perspectives,'' in \emph{9th International Symposium on Computer
  Music Modeling and Retrieval}, vol.~4, 2012.

\bibitem{hoffmann2014music}
P.~Hoffmann, A.~Kaczmarek, P.~Spaleniak, and B.~Kostek, ``Music recommendation
  system,'' \emph{Journal of telecommunications and information technology},
  2014.

\bibitem{downie2003music}
J.~S. Downie, ``Music information retrieval,'' \emph{Annual review of
  information science and technology}, vol.~37, no.~1, pp. 295--340, 2003.

\bibitem{bel1993computational}
B.~Bel and B.~Vecchione, ``Computational musicology,'' \emph{Computers and the
  Humanities}, vol.~27, no.~1, pp. 1--5, 1993.

\bibitem{pfeiffer1997automatic}
S.~Pfeiffer, S.~Fischer, and W.~Effelsberg, ``Automatic audio content
  analysis,'' in \emph{Proceedings of the fourth ACM international conference
  on Multimedia}.\hskip 1em plus 0.5em minus 0.4em\relax ACM, 1997, pp. 21--30.

\bibitem{holland2003preference}
S.~Holland, M.~Ester, and W.~Kie{\ss}ling, ``Preference mining: A novel
  approach on mining user preferences for personalized applications,'' in
  \emph{European Conference on Principles of Data Mining and Knowledge
  Discovery}.\hskip 1em plus 0.5em minus 0.4em\relax Springer, 2003, pp.
  204--216.

\bibitem{dix2009human}
A.~Dix, ``Human-computer interaction,'' in \emph{Encyclopedia of database
  systems}.\hskip 1em plus 0.5em minus 0.4em\relax Springer, 2009, pp.
  1327--1331.

\bibitem{baltrunas2011incarmusic}
L.~Baltrunas, M.~Kaminskas, B.~Ludwig, O.~Moling, F.~Ricci, A.~Aydin, K.-H.
  L{\"u}ke, and R.~Schwaiger, ``Incarmusic: Context-aware music recommendations
  in a car,'' in \emph{International Conference on Electronic Commerce and Web
  Technologies}.\hskip 1em plus 0.5em minus 0.4em\relax Springer, 2011, pp.
  89--100.

\bibitem{wang2018cooperative}
C.~Wang, S.~Gong, A.~Zhou, T.~Li, and S.~Peeta, ``Cooperative adaptive cruise
  control for connected autonomous vehicles by factoring communication-related
  constraints,'' in \emph{The 23rd International Symposium on Transportation
  and Traffic Theory}, 2019.

\bibitem{li2018modeling}
T.~Li, ``Modeling uncertainty in vehicle trajectory prediction in a mixed
  connected and autonomous vehicle environment using deep learning and kernel
  density estimation,'' in \emph{The Fourth Annual Symposium on Transportation
  Informatics}, 2018.

\bibitem{gong2018cooperative}
S.~Gong, A.~Zhou, J.~Wang, T.~Li, and S.~Peeta, ``Cooperative adaptive cruise
  control for a platoon of connected and autonomous vehicles considering
  dynamic information flow topology,'' in \emph{The 21st IEEE International
  Conference on Intelligent Transportation Systems}, 2018.

\bibitem{kaggle2018}
\BIBentryALTinterwordspacing
``Kaggle prediction competition: Music sequence prediction,'' 2018. [Online].
  Available: \url{https://www.kaggle.com/c/music-sequence-prediction}
\BIBentrySTDinterwordspacing

\bibitem{koren2015advances}
Y.~Koren and R.~Bell, ``Advances in collaborative filtering,'' in
  \emph{Recommender systems handbook}.\hskip 1em plus 0.5em minus 0.4em\relax
  Springer, 2015, pp. 77--118.

\bibitem{ekstrand2011collaborative}
M.~D. Ekstrand, J.~T. Riedl, J.~A. Konstan \emph{et~al.}, ``Collaborative
  filtering recommender systems,'' \emph{Foundations and
  Trends{\textregistered} in Human--Computer Interaction}, vol.~4, no.~2, pp.
  81--173, 2011.

\bibitem{gagniuc2017markov}
P.~A. Gagniuc, \emph{Markov Chains: From Theory to Implementation and
  Experimentation}.\hskip 1em plus 0.5em minus 0.4em\relax John Wiley \& Sons,
  2017.

\bibitem{rabiner1989tutorial}
L.~R. Rabiner, ``A tutorial on hidden markov models and selected applications
  in speech recognition,'' \emph{Proceedings of the IEEE}, vol.~77, no.~2, pp.
  257--286, 1989.

\bibitem{hughey1996hidden}
R.~Hughey and A.~Krogh, ``Hidden markov models for sequence analysis: extension
  and analysis of the basic method,'' \emph{Bioinformatics}, vol.~12, no.~2,
  pp. 95--107, 1996.

\bibitem{kan2016map}
\BIBentryALTinterwordspacing
W.~Kan, ``Map@k demo,'' 2016. [Online]. Available:
  \url{https://www.kaggle.com/wendykan/map-k-demo}
\BIBentrySTDinterwordspacing

\bibitem{manning2008chapter}
C.~D. Manning, P.~Raghavan, and H.~Sch{\"u}tze, ``Chapter 8: Evaluation in
  information retrieval,'' \emph{Introduction to information retrieval}, 2008.

\bibitem{li2004similarity}
M.~Li, X.~Chen, X.~Li, B.~Ma, and P.~M. Vit{\'a}nyi, ``The similarity metric,''
  \emph{IEEE transactions on Information Theory}, vol.~50, no.~12, pp.
  3250--3264, 2004.

\bibitem{resnik1999semantic}
P.~Resnik, ``Semantic similarity in a taxonomy: An information-based measure
  and its application to problems of ambiguity in natural language,''
  \emph{Journal of artificial intelligence research}, vol.~11, pp. 95--130,
  1999.

\bibitem{lafferty2001conditional}
J.~Lafferty, A.~McCallum, and F.~C. Pereira, ``Conditional random fields:
  Probabilistic models for segmenting and labeling sequence data,'' 2001.

\bibitem{sutton2006introduction}
C.~Sutton and A.~McCallum, \emph{An introduction to conditional random fields
  for relational learning}.\hskip 1em plus 0.5em minus 0.4em\relax MIT Press,
  2006, vol.~2.

\bibitem{march1991exploration}
J.~G. March, ``Exploration and exploitation in organizational learning,''
  \emph{Organization science}, vol.~2, no.~1, pp. 71--87, 1991.

\bibitem{schumpeter1934theory}
J.~A. Schumpeter, ``The theory of economic development,'' 1934.

\bibitem{gaver2000alternatives}
B.~Gaver and H.~Martin, ``Alternatives: exploring information appliances
  through conceptual design proposals,'' in \emph{Proceedings of the SIGCHI
  conference on Human Factors in Computing Systems}.\hskip 1em plus 0.5em minus
  0.4em\relax ACM, 2000, pp. 209--216.

\bibitem{lecun2015deep}
Y.~LeCun, Y.~Bengio, and G.~Hinton, ``Deep learning,'' \emph{nature}, vol. 521,
  no. 7553, p. 436, 2015.

\bibitem{goodfellow2014generative}
I.~Goodfellow, J.~Pouget-Abadie, M.~Mirza, B.~Xu, D.~Warde-Farley, S.~Ozair,
  A.~Courville, and Y.~Bengio, ``Generative adversarial nets,'' in
  \emph{Advances in neural information processing systems}, 2014, pp.
  2672--2680.

\bibitem{lin2018deep}
L.~Lin, S.~Gong, and T.~Li, ``Deep learning-based human-driven vehicle
  trajectory prediction and its application for platoon control of connected
  and autonomous vehicles,'' in \emph{The Autonomous Vehicles Symposium}, vol.
  2018, 2018.

\bibitem{li2018toward}
T.~Li, K.~Fu, M.~Choi, X.~Liu, and Y.~Chen, ``Toward robust and efficient
  training of generative adversarial networks with bayesian approximation,'' in
  \emph{the Approximation Theory and Machine Learning Conference}, 2018.

\bibitem{rodriguez2017deep}
P.~Rodriguez, G.~Cucurull, J.~Gonzalez, J.~M. Gonfaus, K.~Nasrollahi, T.~B.
  Moeslund, and F.~X. Roca, ``Deep pain: Exploiting long short-term memory
  networks for facial expression classification,'' \emph{IEEE transactions on
  cybernetics}, no.~99, pp. 1--11, 2017.

\bibitem{lopes2017facial}
A.~T. Lopes, E.~de~Aguiar, A.~F. De~Souza, and T.~Oliveira-Santos, ``Facial
  expression recognition with convolutional neural networks: coping with few
  data and the training sample order,'' \emph{Pattern Recognition}, vol.~61,
  pp. 610--628, 2017.

\bibitem{zeng2018facial}
N.~Zeng, H.~Zhang, B.~Song, W.~Liu, Y.~Li, and A.~M. Dobaie, ``Facial
  expression recognition via learning deep sparse autoencoders,''
  \emph{Neurocomputing}, vol. 273, pp. 643--649, 2018.

\bibitem{li2018single}
T.~Li, ``Single image super-resolution: A historical review,'' \emph{ObEN
  Research Seminar}, August 2018.

\bibitem{li2017towards}
X.~Li, H.~Xiaopeng, A.~Moilanen, X.~Huang, T.~Pfister, G.~Zhao, and
  M.~Pietikainen, ``Towards reading hidden emotions: A comparative study of
  spontaneous micro-expression spotting and recognition methods,'' \emph{IEEE
  Transactions on Affective Computing}, 2017.

\bibitem{takalkar2018survey}
M.~Takalkar, M.~Xu, Q.~Wu, and Z.~Chaczko, ``A survey: facial micro-expression
  recognition,'' \emph{Multimedia Tools and Applications}, vol.~77, no.~15, pp.
  19\,301--19\,325, 2018.

\bibitem{dodge1997computer}
C.~Dodge and T.~A. Jerse, \emph{Computer music: synthesis, composition and
  performance}.\hskip 1em plus 0.5em minus 0.4em\relax Macmillan Library
  Reference, 1997.

\bibitem{wang2014guided}
C.-i. Wang and S.~Dubnov, ``Guided music synthesis with variable markov
  oracle,'' in \emph{Tenth Artificial Intelligence and Interactive Digital
  Entertainment Conference}, 2014.

\bibitem{o2017v2hz}
M.~O'Sullivan, B.~Srbinovski, A.~Temko, E.~Popovici, and H.~McCarthy, ``V2hz:
  Music composition from wind turbine energy using a finite-state machine,'' in
  \emph{Signals and Systems Conference (ISSC), 2017 28th Irish}.\hskip 1em plus
  0.5em minus 0.4em\relax IEEE, 2017, pp. 1--6.

\bibitem{colonel2017improving}
J.~Colonel, C.~Curro, and S.~Keene, ``Improving neural net auto encoders for
  music synthesis,'' in \emph{Audio Engineering Society Convention 143}.\hskip
  1em plus 0.5em minus 0.4em\relax Audio Engineering Society, 2017.

\bibitem{ribeiro2018cs}
\BIBentryALTinterwordspacing
B.~Ribeiro, ``Cs57300: Graduate data mining,'' 2018. [Online]. Available:
  \url{https://www.cs.purdue.edu/homes/ribeirob/courses/Spring2018/}
\BIBentrySTDinterwordspacing

\end{thebibliography}
%
%
%

\end{document}